\documentstyle[aps]{revtex}
\newcommand{\dx}{\stackrel{\leftrightarrow}{\partial}_{x}^{\,0}}
\newcommand{\dy}{\stackrel{\leftrightarrow}{\partial}}
\newcommand{\sst}{\scriptscriptstyle}
\newcommand{\be}{\begin{eqnarray}}
\newcommand{\ee}{\end{eqnarray}}
\begin{document}

\title{Boson Interferometry after SPS and before RHIC}
\author{A. Makhlin, E. Surdutovich, and G. Welke$^*$}
\address{Department of Physics and Astronomy,
Wayne State University, Detroit, MI 48202 }
\date{\today}
\maketitle

\begin{abstract}
We re--examine the connection between interferometry and
the Wigner representation for source freeze--out, and
continuous emission.  At the operator level, two equivalent
representations of the two--particle spectrum are found, which
contradict the standard expression of kinetic theory.  The discrepancy is
resolved using two toy models.  Further, we revisit
interferometry in scale--invariant one--dimensional hydrodynamics, and
argue that recent experimental results are evidence for a short kaon
emission time.  Using two exactly calculable models of two-- and
three--dimensional flow, it is shown that the saddle point approximation,
which is reasonable for one--dimensional flow, is no longer adequate.
In these models the scaling law is altered, and we argue that such
qualitative trends, together with other observables, are vital if one
is to draw conclusions about the unknown source parameters.
\end{abstract}


\renewcommand{\theequation}{1.\arabic{equation}}
\setcounter{equation}{0}

\bigskip
\bigskip
\noindent {\large {\bf 1.~Introduction}}
\bigskip

The hope of discovering a quark--gluon plasma (QGP) in heavy ion
collisions is to some extent connected to possibility of measuring the
geometric size of the region of secondary particle production.  A
quantitative estimate of this size is necessary to obtain the energy
density, an important quantity in the discussion of the deconfinement
phase transition.  An important tool in accomplishing such a size
measurement is interferometry.

Recent experiments at the CERN SPS indicate that interferometry is
sensitive to hydrodynamic motion in the emitting source
\cite{NA35,NA44}. At RHIC energies one may expect the signature of
this motion to be even more pronounced. In this paper, we
therefore wish to reexamine various techniques to calculate the
two--boson spectra for expanding sources.

Intensity interferometry was proposed by Hanbury-Brown and Twiss to
measure stellar sizes. The nature of light emission in stars is well
understood; the thermal mechanism guarantees that photons are emitted
independently from different parts of the photosphere, while the small
angular size of the star makes it impossible to obtain an optical
image. These two factors create the conditions necessary for
interference between the two--particle amplitudes of photons emitted
by different parts of the stellar surface. Under certain assumptions
which will be discussed later, the two-particle detection probability
is
\be
W(k_{1},k_{2})\;=\;\int d^{\,4}x_1d^{\,4}x_2\;\rho(x_1)\,\rho(x_2)
\bigg [1\:+\:\cos(k_1-k_2)\!\cdot\!(x_1-x_2)\bigg ] \label{eq:E1.1}\ee

The nature of particle sources in nucleus--nucleus collisions
\cite{review} is less well understood. In particular,
the correspondence between measured quantities and parameters of the
emitting system is less clear. One possible reason Eq.~(\ref{eq:E1.1})
may be inapplicable is the presence of correlations on the same scale
as the size of the emitting system. An extreme case -- when
interferometry is completely unrelated to the total source size -- is
the scattering of coherent light in fluids; the rate of two--photon
coincidence depends on the three-particle distribution function in the
fluid \cite{Mak72}.  For A--A collisions, even in the simplest
scenario of hydrodynamic evolution, the relation between the inclusive
one-- and two--particle spectra and the parameters of the emitting
system does not follow the classical scheme of interferometry. The
width of correlator is not directly connected to the geometric size of
the source, an effect first discussed in Ref.~\cite{Scott}.  For an
expanding spherical shell, the apparent (or ``visible'') source size
is smaller, the larger the total pion pair momentum $P_\perp$. In
nuclear collision models, the dependence of the apparent size on
$P_\perp$ may reflect a more complicated interplay between the
collision dynamics and the true source size \cite{snowb}.

A relativistically covariant theory for interferometry in the
hydrodynamic model of nuclear collisions was developed in
Ref.~\cite{MS}, with the main focus on freeze--out as a realistic
mechanism for final state hadron production.  In a subsequent study
\cite{AMS}, it was shown that if the dependence of the correlation function
on the difference of the longitudinal momenta is rescaled by
$\sqrt{m_\perp}$, the width of the correlator becomes independent of
$m_\perp$.  This scaling behavior has also been discovered in parallel
studies \cite{KG} based on the intra-nuclear cascade approach
\cite{GKW}, where classical currents are the sources of final--state
pions.  The $m_\perp$ scaling dependence has been observed at SPS
energies by the NA35 and NA44 collaborations \cite{NA35,NA44}.

A covariant theory of interference for continuous emission has also
been applied to dilepton \cite{dilep} and photon \cite{Phot,Kap}
emission from a quark--gluon plasma.  Analysis of polarization
effects in photon interference \cite{Phot} indicates that there is a
dilemma in the proper choice of formalism: the languages of
locally defined states, and Wigner phase-space distributions
lead to different answers.

Events generators such as RQMD \cite{RQMD} suggest a classical
description of the space--time evolution of high energy heavy ion
collisions.  Other models, based on the solution of semi--classical
kinetic equations, often provide descriptions in terms of
one--particle distributions. Interferometry is a manifestly
quantum phenomenon, and we shall therefore begin in Section~2 by
examining its connection to the widely used Wigner representation, for
the cases of source freeze--out (initial data problem) and photons
(continuous emission).  At the operator level, the Wigner
representation is simply a formal re--expression of the one-- and
two--particle spectra, and we obtain two equivalent expressions for
the two--particle spectrum.  These are in contradiction with the
standard expressions of kinetic theory.  In Section~3, we attempt to
resolve the issue by way of two toy models, {\it viz.}, the emission of
particles from one and two cavities. In an appendix, we reanalyze the
derivation of the two-particle spectrum in classical source models.

In Section~4, we shall revisit interferometry in scale--invariant
one--dimensional hydrodynamics, deriving the $m_\perp$--dependence of
the longitudinal effective size in the case of the ``extended
freeze--out.''  We argue that recent experimental results \cite{NA44}
confirming this behavior are evidence for short kaon emission times,
and in disagreement with the prediction of kaon distillation \cite{destil}.

Unlike the inverse scattering problem, where an analysis of the phase
shifts allows one to obtain the shape of the potential, the inverse
problem of interferometry has no exact mathematical formulation. The
interpretation of the two--particle spectra requires a model that is
determined {\it a priori}, up to the numerical values of its
parameters.  We therefore present in Section~5 some results for two--
and three--dimensional flow in exactly calculable examples: the
transverse explosion of a long filament, and the spherical explosion
of a point--like source. In both cases the saddle point approximation,
which is reasonable for one--dimensional flow, is no longer adequate.
The scaling law is altered, and we argue that such qualitative trends,
together with other observables, are important in establishing the
type of model one is dealing with.  Only once the model is given, can
one draw conclusions about the unknown source parameters. We conclude
in Section~6.

\renewcommand{\theequation}{2.\arabic{equation}}
\setcounter{equation}{0}

\bigskip
\bigskip
\noindent {\large {\bf 2.~Initial data and continuous
emission for interferometry}}
\bigskip


The collision dynamics in heavy ion physics is rather complicated.
However, some stages of the collision may be described
in terms of one--particle distributions, or even by a few macroscopic
parameters.  A strong, yet attractive, simplification occurs
if we assume that the hot expanding matter freezes out at some
critical temperature $T_c$. Generally, such models rely
on (semi--) classical assumptions, and produce (semi--) classical final
distributions. On the other hand, interferometry is a manifestation of
the quantum nature of the constituents of the system, and it is
therefore necessary to examine carefully whether these distributions
can serve as an adequate input.

A natural connection between quantum mechanics and (semi--) classical
descriptions is provided by the density matrix $\rho$.  In a formal
language, the problem we wish to solve in interferometry is to
determine judiciously chosen parameters of $\rho$ by measuring the
inclusive cross-sections $dN_{1}/d{\vec k}$, $dN_{2}/d{\vec k}_{1}
d{\vec k}_{2},\ldots$~.

\bigskip
\noindent {2.1 \underline{\it Source freeze--out}}
\bigskip

Let $|{\rm in}\rangle$ be one of the possible initial states of the
system emitting a pion field ${\hat \varphi}(x)$.  At this stage,
we do not wish to specify the exact nature of the initial
state, but study instead the observables relevant for
interferometry at the operator level.

The field ${\hat \varphi}(x)$ is to be detected later by some device,
or analyzer.  Let it be tuned to the measurement of the momentum of
the free pion. Thus, the eigenfunctions of the analyzer are the free
pion wave functions
\be
f_{\vec k}(x)\;=\;(2\pi)^{-3/2}(2k_{0})^{-1/2}
\:e^{-i\,k\cdot x}~ ~ ~.\label{eq:E2.1}
\ee
These functions are defined for $x^0>t_c$, where $t_c$ is
the time when the system ``prepared'' the final pion.
The corresponding annihilation operator $A_{\vec k}$ for
momentum ${\vec k}$ in the final state is given by
\be
{\hat A}_{\vec k}\;=\;\int d^{3}x\; f^{*}_{\vec k}(x)\,i\!\dx
{\hat \varphi}(x)~ ~ ~, \label{eq:E2.2}
\ee
where $(a\!\!\dx\!\!b)\equiv a(\partial_x^{\,0}\,b) -
(\partial_x^{\,0}\,a)b$.  The operator ${\hat A}_{\vec k}$ describes
the effect of a detector (analyzer) far from the point of emission,
so, by definition, the pion is detected on mass--shell, $k^0=({\vec
k}^2 +m^2)^{1/2}$.  The inclusive amplitude to find one pion with
momentum ${\vec k}$ in the final state is
\be
\langle X|{\hat A}_{\vec k}\;{\hat S}\,|{\rm in}\rangle~ ~ ~,
\label{eq:E2.3}
\ee
where ${\hat S}$ is the evolution operator after freeze--out, and
the states $|X\rangle$ form a complete set of all possible
secondaries. Summing the squared modulus of this amplitude over all
(undetected) states $|X\rangle$, and averaging over the initial
ensemble, we find the one--particle inclusive spectrum
\be
{ dN_{1} \over d{\vec k} } \; = \; {\rm Tr}\; {\hat \rho}_{\rm in}\,
{\hat S}^\dagger {\hat A}^\dagger_{\vec k} {\hat A}_{\vec k}\,
{\hat S}~ ~ ~,\label{eq:E2.4}
\ee
where the density operator ${\hat \rho}_{\rm in}$ will describe the
emitting system.  For now, we shall study only the operator part of
Eq.~(\ref{eq:E2.4}), ${\hat N}_{\vec k} = {\hat A}^\dagger_{\vec
k}{\hat A}_{\vec k}$, which gives the number of pions detected by an
analyzer tuned to momentum $\vec k$.  In the same way one may obtain
the inclusive two--pion spectrum
\be
{  {dN_{2}} \over {d{\vec k}_{1} d{\vec k}_{2} } }
\; =\; {\rm Tr}\; {\hat \rho}_{\rm in}\, {\hat S}^\dagger
{\hat A}^\dagger_{{\vec k}_{1}}{\hat A}^\dagger_{{\vec k}_{2}}
{\hat A}_{{\vec k}_{2}}{\hat A}_{{\vec k}_{1}}\,{\hat S}~ ~ ~,
\label{eq:E2.7}
\ee
and the ``device'' operator which counts the number of pions pairs is
\be
{\hat A}^\dagger_{{\vec k}_{1}} {\hat A}^\dagger_{{\vec k}_{2}}
{\hat A}_{{\vec k}_{2}} {\hat A}_{{\vec k}_{1}}\;=\;
{\hat N}_{{\vec k}_1} ({\hat N}_{{\vec k}_2}-
\delta({\vec k}_1-{\vec k}_2) )~ ~ ~.\label{eq:E2.8}
\ee

The pion field ${\hat \varphi}(x)$ reaches the detector after free
propagation.  This simplest kind of evolution is described by the
retarded Green function
\be
{\hat \varphi}(x) &=& \int d\Sigma_{\mu}(y)\:G_{\rm ret}(x-y)\dy^{\mu}_{y}
{\hat \varphi}(y), \nonumber \\ {\hat \varphi}^\dagger (x)
&=& \int d\Sigma_{\mu}(y)\:{\hat \varphi}^\dagger (y)\dy^{\mu}_{y} G_{\rm
adv}(y-x) ~ ~ ~,\label{eq:E2.9}
\ee
where the pion field ${\hat \varphi}(y)$ is given on the 3-dimensional
freeze--out hypersurface (a Cauchy hypersurface).  The space of states
in which the density matrix acts is also defined on this surface.
Substituting Eqs.~(\ref{eq:E2.9}) in (\ref{eq:E2.2}), we find the
final form of the pion Fock operator, expressed in terms of the
initial fields:
\be
{\hat S}^\dagger {\hat A}_{\vec k}\,{\hat S} &=&
\int d^{\,3}x\: f^{*}_{\vec k}(x)\,i\!\dx
\int d\Sigma_{\mu}(y)\:G_{\rm ret}(x-y)\dy^{\mu}_{y}
{\hat \varphi} (y)~ ~ ~,\nonumber \\
{\hat S}^\dagger {\hat A}^\dagger_{\vec k}\,{\hat S} &=&\int d^{\,3}x
\int d\Sigma_{\mu}(y)\: {\hat \varphi}^\dagger
(y)\dy^{\mu}_{y}G_{\rm adv}(y-x)\, i\!\dx f_{\vec k}(x)~ ~ ~.
\label{eq:E2.10}
\ee
These equations may be simplified using the explicit form
of the  free pion propagators:
\be
G_{\stackrel {\rm ret}{\rm {\sst adv}}}(x-y)\;=\;
\int {d^4p \over (2\pi)^4} \; {e^{-i\,p\cdot (x-y)} \over
(p^0\pm i\epsilon)^2-{\vec p}^{\,2}-m^2 }~ ~ ~.\label{eq:E2.11}
\ee
Substituting Eqs.~(\ref{eq:E2.11}) into (\ref{eq:E2.10}) and
integrating, first over $d^{\, 3}x$ which sets ${\vec p}$ equal to
${\vec k}$, and then over $p^0$, we obtain
\be
{\hat S}^\dagger {\hat A}_{\vec k}\,{\hat S}
&=& \theta(x^0-y^0)\int d\Sigma_{\mu}(y)
\; f^*_{\vec k}(y) i\!\dy^{\mu}_{y} {\hat \varphi} (y)~ ~ ~, \nonumber\\
{\hat S}^\dagger {\hat A}^\dagger_{\vec k}\,{\hat S}
&=& \theta(x^0-y^0)\int d\Sigma_{\mu}(y)\; {\hat \varphi}^\dagger (y)
i\!\dy^{\mu}_{y} f_{\vec k}(y)~ ~ ~.\label{eq:E2.12}
\ee
As before, the pion four--momentum $k$ is on mass--shell.
This is not surprising, as the propagation is free and our analyzers perform
an on--shell Fourier expansion of the initial data. The
$\theta$--function appears because the propagation is retarded.
Using (\ref{eq:E2.12}), we may rewrite the number operators for single
pions and pairs of pions as
\be
{\hat N}_{\vec k}\:=\:\theta(x^0\!-\!y^0)\int\! d\Sigma_{\mu}(y_1)\,
d\Sigma_{\nu}(y_2)\:\bigg [ f_{\vec
k}(y_1)\,i\!\dy^{\,\mu}_{y_1}{\hat \varphi}^\dagger
(y_1)\bigg ]\:\bigg [ {\hat \varphi} (y_2)\, i\!
\dy^{\,\nu}_{y_2} f^{*}_{\vec k}(y_2)\bigg ]~ ~ ~\label{eq:E2.13}
\ee
and
\be
{\hat N}_{{\vec k}_1}({\hat N}_{{\vec k}_2}-\delta({\vec k}_1-{\vec k}_2) )
\:=\:\theta(x^0-y^0)\int d\Sigma_{\mu}(y_1)\,d\Sigma_{\nu}(y_3)\,
d\Sigma_{\rho}(y_4)\,d\Sigma_{\lambda}(y_2)  \hspace*{2cm} \nonumber\\
\times\, \bigg [f_{{\vec k}_1}(y_1)f_{{\vec k}_2}(y_3) \dy^{\,\mu}_{y_1}
\dy^{\,\nu}_{y_3}
{\hat \varphi}^\dagger (y_1){\hat \varphi}^\dagger (y_3) \bigg ]\:
\bigg [ {\hat \varphi}
(y_4){\hat \varphi} (y_2)\dy^{\,\rho}_{y_4}\dy^{\,\lambda}_{y_2}
f^{*}_{{\vec k}_2}(y_4) f^{*}_{{\vec k}_1}(y_2)\bigg ]~,
\hspace*{0.6cm}\label{eq:E2.14}
\ee
respectively.

We shall now attempt to incorporate the technique used in kinetic
studies by passing to the Wigner representation at the operator level,
{\it i.e.}, without specifying any physical information about emitting
system. Firstly, for ${\hat N}_{\vec k}$, we choose new variables
\be
{\vec y}_1={\vec R} +{{\vec z} \over 2}~ ~,\;\;\;~ ~ ~
{\vec y}_2={\vec R} -{{\vec z} \over 2} ~ ~ ~.\label{eq:E2.15}
\ee
This transformation is motivated by a simple idea: we expect that
${\vec R}$ will label the sources, while the Fourier transform over
${\vec z}$ will yield the local spectrum.  We assume that the
3--vectors ${\vec R}$ and ${\vec z}$ lie in the Cauchy hypersurface,
which is obviously not planar, in general.  The local time--like
direction coincides with the normal.  As the equation for ${\hat
\varphi}$ is second order, the initial data includes the derivatives
$\dy_{y_i}^{0}$, which lie along the local normal.  To adjust the
variables, we write
\be
\dy_{y_1}^{\,0}\;=\;{1\over2}\!\dy_R^{\,0}\:+\:\dy_z^{\,0}~ ,\;\;\;~ ~ ~ ~
\dy_{y_2}^{\,0}\;=\;{1\over2}\!\dy_{R}^{\,0}\:-\:\dy_{z}^{\,0}~ ~ ~,
\label{eq:E2.21}
\ee
allowing us to introduce a compact form of the Wigner representation
of the operator (\ref{eq:E2.13})
\be
 {\hat {\cal N}}(R,{k_1 +k_2 \over 2}) \;=\;
\int d^{\, 3}z\: e^{-i({\vec k}_1 +{\vec k}_2)\cdot{\vec
z}/2} \: e^{-i\,k_1^0\cdot (R^0+z^0/2)}\:i \bigg [
{1\over2}\dy_{R}^{0}+\dy_{z}^{0}\bigg ] \: {\hat \varphi}^\dagger
(R^0\!+\!\frac{z^0}{2},{\vec R}\!+\!\frac{\vec z}{2} )
\nonumber \\ \hspace*{0.6cm} \times \:
{\hat \varphi}(R^0\!-\!\frac{z^0}{2},{\vec R}\!-\!\frac{\vec z}{2} )
\: i \bigg [ {1\over2}\dy_R^0-\dy_z^0 \bigg ] \:
e^{i\,k_2^0\cdot (R^0+z^0/2)}~ ~ ~,\label{eq:E2.22}
\ee
which may be shown to depend only on the combination
$(k_{1}^{0}+k_{2}^{0})/2$.  If the Cauchy surface is planar, all the
derivatives above will be along the common normal. Then, the one
particle spectrum may be represented in a compact symbolic form
\be
{\hat N}_{\vec k}\;=\;{1\over  2 k^0}\int\! d^{\,3}R \;
{\hat {\cal N}}(R, k)~ ~ ~. \label{eq:E2.23}
\ee

For the two--particle operator we have four arguments in the integrand
of (\ref{eq:E2.14}), which allows us to choose the variables for the
Wigner transformation in two ways:
\be
{\vec y}_1&=&{\vec R}_1 +{{\vec z}_1 \over 2},\;\;\;
{\vec y}_2\;=\;{\vec R}_1 -{{\vec z}_1 \over 2}, \nonumber \\
{\vec y}_3&=&{\vec R}_2 +{{\vec z}_2 \over 2},\;\;\;
{\vec y}_4\;=\;{\vec R}_2 -{{\vec z}_2 \over 2}~ ~ ~;\label{eq:E2.17}
\ee
or, equally,
\be
{\vec y}_1&=&{\vec R}_1 +{{\vec z}_1 \over 2},\;\;\;
{\vec y}_4\;=\;{\vec R}_1 -{{\vec z}_1 \over 2}, \nonumber \\
{\vec y}_3&=&{\vec R}_2 +{{\vec z}_2 \over 2},\;\;\;
{\vec y}_2\;=\;{\vec R}_2 -{{\vec z}_2 \over 2} ~ ~ ~.\label{eq:E2.19}
\ee
Correspondingly, the operators for the two--particle spectrum may be rewritten
in the two identical forms
\be
{\hat N}_{{\vec k}_1}({\hat N}_{{\vec k}_2}-\delta({\vec k}_1-{\vec
k}_2))\;=\;{1\over  4 k_{1}^{0}k_{2}^{0}} \int \!d^{\,3}R_1\, d^{\, 3}R_2
\; {\hat {\cal N}}(R_1,k_1) \:
{\hat {\cal N}}(R_2, k_2) \label{eq:E2.24a}
\ee
and
\be
{\hat N}_{{\vec k}_1}({\hat N}_{{\vec k}_2}-\delta({\vec k}_1-{\vec
k}_2) ) \;=\;
{1\over  4 k_{1}^{0}k_{2}^{0}} \int \! d^{\, 3}R_1 d^{\, 3}R_2
\;{\hat {\cal N}}(R_1,{k_1 + k_2 \over 2})\:
{\hat {\cal N}}(R_2,{k_1 + k_2 \over 2})\;
e^{-i\,({\vec k}_1 -{\vec k}_2)\cdot ({\vec R}_1 -{\vec
R}_2)}~ ~,\label{eq:E2.24}
\ee
respectively.

\bigskip
\noindent {2.2 \underline{\it Continuous emission}}
\bigskip

We shall now consider continuous emission from an evolving system,
using the case of photons as an example \cite{Phot}.
Unlike for pion emission, where the (transition) currents which
might  emit the pions are not conserved, this problem has a clearer
footing. Also, technically, the Wigner representation for continuous
emission is simpler, as there are no derivatives in the equations
that will replace (\ref{eq:E2.13}) and (\ref{eq:E2.14}). As in the
previous section,
we define the inclusive probabilities for one and two photons
\be
{{dN_{\gamma}} \over {d{\vec k}}} \;=\; \sum_{\lambda} \: {\rm Tr}\:
{\hat \rho}_{\rm in} \: {\hat S}^\dagger\, {\hat c}^\dagger({\vec k},\lambda)
{\hat c}({\vec k},\lambda)\,{\hat S}~ ~ ~,\label{eq:2.25}
\ee
and
\be
{dN_\gamma \over d{\vec k}_1 d{\vec k}_2} \;=\;\sum_{\lambda_1,\lambda_2}\:
{\rm Tr}\:{\hat \rho}_{\rm in}\:
{\hat S}^\dagger\, {\hat c}^\dagger ({\vec k}_1,\lambda_1)
{\hat c}^\dagger ({\vec k}_2,\lambda_2) {\hat c}({\vec k}_2,\lambda_2)
{\hat c}({\vec k}_1,\lambda_1)\,{\hat S}~ ~ ~,\label{eq:E2.26}\ee
respectively. The operators ${\hat c}({\vec k})$ appear in the
decomposition of the electromagnetic field operator,
\be
{\hat A}_{\mu}(x)= \sum_{\lambda} \int d^{\, 3}k\:
{\epsilon^{(\lambda)}_{\mu}({\vec k}) \over (2\pi)^{3/2}(2
\omega_{\vec k})^{1/2}}
\;\bigg [{\hat c}({\vec k},\lambda)e^{-i\,k\cdot x}+{\hat c}^\dagger ({\vec
k},\lambda)e^{i\,k\cdot x} \bigg ]~ ~ ~,\label{eq:E2.27}\ee
where $\lambda$ runs over the two physical polarizations of the photon.
Since the quantum equation of motion for the field
${\hat A}_{\mu}(x)$ has the classical form,
\be
{\hat A}^{\mu}(x) \;=\;  \int d^{\,4}y\; D_{ret}^{\mu\nu}(x-y)\:{\hat
j}_{\nu}(y)~ ~ ~,\label{eq:E2.28}\ee
it is easy to show that the operator of the number of emitted photons is
\be
 {\hat N}_{\vec k} = \sum_{\lambda} \epsilon^{\mu,(\lambda)}({\vec k})
\epsilon^{\nu,(\lambda)}({\vec k})
\int {d^{\, 4}xd^{\, 4}y\over 2k^0 (2\pi)^3}\; e^{-i\,k\cdot (x-y)}
\;{\hat j}^\dagger_{\mu}(x)\:{\hat j}_{\nu}(y)~ ~ ~.\label{eq:E2.29}\ee
Now we may safely perform a full 4--dimensional Wigner transformation
\be
x^\mu=R^\mu +{z^\mu \over 2}~,~ ~ ~\;\;\; y^\mu=R^\mu -{z^\mu \over 2}
{}~ ~ ~.\label{eq:E2.31}\ee
Introducing
\be
{\hat \Pi}_{\mu\nu}(R,p)\;=\; \frac {1}{2k^0} \:
\int {d^{\,4}z\over (2\pi)^4}\; e^{-i\,p\cdot z}\;
{\hat j}^\dagger_{\mu}(R+z/2)\:{\hat j}_{\nu}(R-z/2)~ ~ ~,\label{eq:E2.32}\ee
we may rewrite
\be
 {\hat N}_{\vec k}\;=\; \sum_{\lambda} \epsilon^{(\lambda)}_{\mu}({\vec k})
\epsilon^{(\lambda)}_{\nu}({\vec k})
\int d^{\, 4}R\; {\hat \Pi}^{\mu\nu}(R,k)~ ~ ~.\label{eq:E2.33}\ee
For the operator that counts the number of photon pairs,
\be
{\hat N}_{{\vec k}_1}({\hat N}_{{\vec k}_2}-\delta({\vec k}_1-{\vec
k}_2) ) \;=\;
\sum_{\lambda_1,\lambda_2} \epsilon^{\mu,(\lambda_1)}({\vec k}_1)
\epsilon^{\rho,(\lambda_1)}({\vec k}_1)
\epsilon^{\nu,(\lambda_2)}({\vec k}_2)
\epsilon^{\sigma,(\lambda_2)}({\vec k}_2) \hspace*{2cm}\nonumber\\
\times \;\int {d^{\,4}x\, d^{\,4}y\, d^{\,4}x^\prime\, d^{\,4}y^\prime
\over 4k_{1}^{0}k_{2}^{0} (2\pi)^6} e^{-i\,k_1\cdot
(x-y)-i\,k_2\cdot (x^\prime-y^\prime)}
\;{\hat j}^\dagger_{\mu}(x)\,{\hat j}^\dagger_{\nu}(x^\prime)\,
{\hat j}_{\rho}(y)\,{\hat j}_{\sigma}(y^\prime)~ ~ ~,\hspace*{0.8cm}
\label{eq:E2.30}\ee
The two--photon detection function can be rewritten two ways, by
changing variables either as
\be
x^\mu=R_{1}^{\mu} +{z_{1}^{\mu} \over 2}~,\;\;\;~ ~
y^\mu=R_{1}^{\mu} -{z_{1}^{\mu} \over 2}~,\;\;\;~ ~
x^{\prime \,\mu}=R_{2}^{\mu} +{z_{2}^{\mu} \over 2},\;\;\;~ ~
y^{\prime \,\mu}=R_{2}^{\mu} -{z_{2}^{\mu} \over 2}~,
\label{eq:E2.34}\ee
or,
\be
x^\mu=R_{1}^{\mu} +{z_{1}^{\mu} \over 2}~,\;\;\;~ ~
y^{\prime\,\mu}=R_{1}^{\mu} -{z_{1}^{\mu} \over 2}~,\;\;\;~ ~
x^{\prime \,\mu}=R_{2}^{\mu} +{z_{2}^{\mu} \over 2}~,\;\;\;~ ~
y^\mu=R_{2}^{\mu} -{z_{2}^{\mu} \over 2}~.\label{eq:E2.36}\ee
These transformations lead to two equivalent representations of the
two--photon spectrum:
\be
{\hat N}_{{\vec k}_1} ({\hat N}_{{\vec k}_2}-\delta({\vec k}_1-{\vec k}_2) )
\;=\;\sum_{\lambda_1,\lambda_2} \epsilon^{(\lambda_1)}_{\mu}({\vec k}_1)
\epsilon^{(\lambda_1)}_{\rho}({\vec k}_1) \epsilon^{(\lambda_2)}_{\nu}
({\vec k}_2)
\epsilon^{(\lambda_2)}_{\sigma}({\vec k}_2)
\;\int d^{\,4}R_1 d^{\,4}R_2 \;{\hat \Pi}^{\mu\rho}(R_1,k_1)
\:{\hat \Pi}^{\nu\sigma}(R_2,k_2)\label{eq:E2.35}\ee
and
\be
{\hat N}_{{\vec k}_1} ({\hat N}_{{\vec k}_2}-\delta({\vec k}_1-
{\vec k}_2) )\;=\; \sum_{\lambda_1,\lambda_2}
\epsilon^{(\lambda_1)}_{\mu}({\vec k}_1)
\epsilon^{(\lambda_1)}_{\rho}({\vec k}_1)
\epsilon^{(\lambda_2)}_{\nu}({\vec k}_2)
\epsilon^{(\lambda_2)}_{\sigma}({\vec k}_2) \hspace*{2cm}\nonumber \\
\times\; \int d^{\,4}R_1 d^{\,4}R_2 \;
{\hat \Pi}^{\mu\nu}(R_1,{k_1+k_2 \over 2})\:
{\hat \Pi}^{\rho\sigma}(R_2,{k_1+k_2 \over 2})\:
\cos(R_1\!-\!R_2)\!\cdot\!(k_1\!-\!k_2) ~.\label{eq:E2.37}\ee
respectively.

\bigskip
\noindent {2.3 \underline{\it Discussion}}
\bigskip

We have obtained three
different expressions for the two-particle spectra, for both
the initial data (freeze--out)
problem and for continuous emission, {\it viz},
Eqs.~(\ref{eq:E2.14}), (\ref{eq:E2.24a}) and (\ref{eq:E2.24}), and
(\ref{eq:E2.30}), (\ref{eq:E2.35}) and (\ref{eq:E2.37}),
respectively. An apparent paradox appears
if one compares these expressions with those commonly used.
For the freeze--out mechanism of the pion production these expressions
have the form
\be
4 k_1^0k_2^0 {dN_{\pi}\over d{\vec k}_1 d{\vec k}_2} &=&
 \int d^{\,3}R_1 d^{\,3}R_2 \;{\cal N}({\vec R}_1,k_1)
\:{\cal N}({\vec R}_2, k_2) \;+\;\hspace*{4cm}    \label{eq:E2.38} \\
&+&\int d^{\, 3}R_1 d^{\,3}R_2\;{\cal N}({\vec R}_1,{k_1 + k_2 \over 2})
\:{\cal N}({\vec R}_2,{k_1 + k_2 \over 2})\; e^{-i\,({\vec k}_1
-{\vec k}_2)\cdot ({\vec R}_1-{\vec R}_2)}~,\nonumber\ee
while for continuous photon emission
\be
4 k_1^0k_2^0 {dN_{\gamma}\over d{\vec k}_1 d{\vec k}_2} &=&
\sum_{\lambda_1,\lambda_2} \epsilon^{(\lambda_1)}_{\mu}({\vec k}_1)
\epsilon^{(\lambda_1)}_{\rho}({\vec k}_1) \epsilon^{(\lambda_2)}_{\nu}
({\vec k}_2)
\epsilon^{(\lambda_2)}_{\sigma}({\vec k}_2)
\int d^{\,4}R_1 d^{\,4}R_2 \times\label{eq:E2.39} \\
&\times& \bigg [ \Pi^{\mu\rho}(R_1,k_1)\Pi^{\nu\sigma}(R_2,k_2)\: +\:
 e^{-i\,(R_1-R_2)\cdot (k_1-k_2)} \Pi^{\mu\nu}(R_1,{k_1+k_2 \over 2})
\Pi^{\rho\sigma}(R_2,{k_1+k_2 \over 2}) \bigg ]~. \nonumber\ee
(For references and a detailed derivation of such formulae, see
Ref.~\cite{zajc}; there, the off--mass--shellness of the distributions
in the cross--term is emphasized as a physical result.)  These
expressions appear to be in line with kinetic theory: they seem to
allow one to use the (semi--) classical phase-space distributions as
the input for subsequent studies of the interferometry problem. Below,
we shall resolve the manifest contradiction between
Eqs.~(\ref{eq:E2.38}) and (\ref{eq:E2.39}) on the one hand, and the
expressions given in the previous sections on the other. For now, we
emphasize that the phase--space densities with argument $(k_1+k_2)/2$
cannot be considered as distributions of the single particles in
definite quantum states.  For the photons, there appears an additional
discrepancy between the transversality of the polarization tensor, and
the physical polarization of the detected photons \cite{Phot}.

\renewcommand{\theequation}{3.\arabic{equation}}
\setcounter{equation}{0}

\bigskip
\bigskip
\noindent {\large {\bf 3.~Interferometry}}
\bigskip

Up to this point, we have only discussed how the particles are
detected, and have yet to describe the states in which particles were
created.  We have seen that the formal Wigner transformation puts the
one--particle operators into a form resembling classical phase--space
distributions, but that the representation for the two--particle
``detection'' operators is not uniquely defined. It allows for two
{\it equivalent} forms, each of which, by coincidence, resembles one
of the terms of the familiar interferometric formula,
Eq.~(\ref{eq:E2.38}), or Eq.~(\ref{eq:E2.39}).  Each Wigner operator
representation is supposed to yield the two term interferometric
formula as a final result.

To proceed with a discussion of interferometry, we require physical
input. This input is already present in equations like (\ref{eq:E2.4})
and (\ref{eq:E2.7}), where the operators are averaged with the density
matrix of the initial state.  It is not evident that any arbitrary
input will make interferometry which yields
physically useful information possible.

Interference of quantum mechanical amplitudes appears when for
identical initial states and identical final states (of one or more
particles) there exists more than one intermediate transition
amplitude.  Interference in one--particle propagation exists, and is
used in optical devices to obtain an optical image of the source.
Interference in the propagation of the two--particle amplitude does
not necessarily exist, because there may not be two alternative paths
for the evolution of the two--particle state.  An example is
two--photon emission from a single atomic transition, and for each
different case, we must examine the initial state of the
system, {\it i.e.}, the density matrix $\rho_{\rm in}$.

\bigskip
\bigskip
\noindent {3.1 \underline{\it
Toy model 1. Emission from a one--dimensional cavity}}
\bigskip

Consider the case of an initial state which is a one--dimensional
cavity filled with hot bosonic radiation.  For the sake of simplicity,
we shall impose periodic boundary conditions on the walls of the
cavity. The cavity eigenfunctions are
\be
\phi_{p}(y)&=&(2a)^{-1/2}(2p^{0})^{-1/2} e^{-i\,p\cdot y}~, ~
{}~ ~ ~ {\rm for}~ -a\le y\le a~,\nonumber\\
&=& 0~, ~ ~ ~ ~{\rm otherwise} \label{eq:E3.1}
\ee
where $2a$ is the distance between the walls, and the allowed values of
the momentum $p$ are
\be
 p_n={n \pi \over a},~~~ n=0,\pm 1, \pm 2,...~ ~ ~.\label{eq:E3.2}\ee
The  boson annihilation and creation operators in the cavity are
defined by their field decomposition:
\be
{\hat\varphi}(y)= \sum_p {\hat a}_{p}\:\phi_{p}(y)~ ~ ~,\label{eq:E3.3}\ee
so that, if the walls are removed at time $y^0=0$,
the average number of particles with the momentum k is given by
(see Eq.~(\ref{eq:E2.13}))
\be
\langle {\hat N}_k\rangle \;=\; \int_{-a}^{a}dy_1 dy_2\: \sum_{p}n(p)\;
\bigg [ f_k(y_1)\dy^{\, 0}_{y_1}\phi^{*}_{p}(y_1)\bigg ]\;\bigg [
\phi_{p}(y_2) \dy^{\, 0}_{y_2}  f^{*}_k(y_2)\bigg ]~ ~ ~,\label{eq:E3.6}\ee
where $n(p)= \langle
{\hat a}^\dagger_{p}{\hat a}_{p}\rangle$ is the boson occupation number.
The integrals may be evaluated to yield
\be
\langle {\hat N}_k\rangle\;=\;\sum_{p} \: n(p)  \:
{(\omega_k + \omega_p)^2 \over 4\omega_k \omega_p }\;
{\sin^2 (p-k)a \over \pi a (p-k)^2  }~ ~ ~.\label{eq:E3.7}\ee

Using Eq.~(\ref{eq:E2.14}), we may similarly write for
the two--particle spectrum
\be
\langle {\hat N}_{k_1}({\hat N}_{k_2}-\delta_{k_1k_2})\rangle \;=\;
\sum_{p_1,..,p_4} {(\omega_{k_1}\!+\!\omega_{p_1})
(\omega_{k_1}\!+\!\omega_{p_2})
(\omega_{k_2}\!+\!\omega_{p_3})(\omega_{k_2}\!+\!\omega_{p_4})
 \over 4\pi^2 4a^2 4 \omega_{k_1} \omega_{k_2}
\sqrt{\omega_{p_1}\omega_{p_2}\omega_{p_3}\omega_{p_4}} }
\hspace{2cm}\nonumber \\
\times \;\langle {\hat a}^\dagger_{p_1} {\hat a}^\dagger_{p_3}
{\hat a}_{p_4}{\hat a}_{p_2} \rangle \;
\int_{-a}^{a}\!\!dy\, e^{-i\,(k_1-p_1)y}
\int_{-a}^{a}\!\!dy\, e^{i\,(k_1-p_2)y}
\int_{-a}^{a}\!\!dy\, e^{-i\,(k_2-p_3)y}
\int_{-a}^{a}\!\!dy\, e^{i\,(k_2-p_4)y}~.\nonumber \\ \label{eq:E3.8}\ee
In case of the Gibbs ensemble, the average of the four Fock operators
is easily calculated
\be
\langle a^\dagger_{p_1} a^\dagger_{p_3}a_{p_4}a_{p_2} \rangle
\;=\;n(p_1)n(p_3)\;\bigg [\delta_{p_1p_2}\delta_{p_3 p_4} +
\delta_{p_1p_4}\delta_{p_3 p_2} \bigg ] ~ ~ ~,\label{eq:E3.9}\ee
so that
\be
\langle {\hat N}_{k_1}({\hat N}_{k_2}-\delta_{k_1k_2})\rangle \;=\;
\langle {\hat N}_{k_1}\rangle\: \langle {\hat N}_{k_2}\rangle
&+&\sum_{p_1,p_3} n(p_1)n(p_3)\;
{(\omega_{k_1}\! +\! \omega_{p_1})(\omega_{k_1}\! +\! \omega_{p_3})
(\omega_{k_2}\! +\! \omega_{p_1})(\omega_{k_2}\! +\! \omega_{p_3})
\over 4\pi^2 4a^2 4 \omega_{k_1} \omega_{k_2}
\omega_{p_1}\omega_{p_3} } \nonumber \\
&\times&{\sin (p_1-k_1)a \over (p_1-k_1)}\:{\sin (p_1-k_2)a \over (p_1-k_2)}
\:{\sin (p_3-k_1)a \over (p_3-k_1)}\:{\sin (p_3-k_2)a \over (p_3-k_2)}
\label{eq:E3.10}\ee

Let us first discuss the single particle spectrum, (\ref{eq:E3.7}).
{}From Eq.~(\ref{eq:E3.2}), it is clear that $\langle {\hat N}_k\rangle$
has an infinite sequence of alternating zeros and peaks, in analogy to
an optical Fabry--Perot interferometer, where the oscillations arise
from the first order (one-particle) multi-ray interference
\cite{Born}. Here, we rely on the interference between all possible
histories of a single particle, and, in principle, the distance
between two neighboring zeros of $\langle {\hat N}_k\rangle$ will
determine the size of the cavity.

The two--particle spectrum, Eq.~(\ref{eq:E3.10}), has a rather complex
structure. It arises because any given two--particle final state can
be traced back to an infinite number of initial states; when the
cavity is opened, every eigenstate of the finite--sized cavity has, in
addition to its ``fundamental harmonic,'' an infinite sequence
secondary peaks, or ``satellites.'' A given plane wave particle may
thus have originated from any single--particle level in the cavity.
The resulting unwanted structure is an artifact of our choice of
boundary conditions, and would dissapear if the zeros in $\langle
{\hat N}_k\rangle$ were not equidistant.

As long as the detailed behavior inherent in this model is not
expected to appear in a real physical application of interferometry,
it is reasonable to modify it by eliminating the rapid oscillations in
the two--particle spectrum Eq.(\ref{eq:E3.10}). As artificial as the
model is, any procedure will do. {\it Ad hoc}, one could remove the
secondary spectral lines by hand, restricting the sum in
(\ref{eq:E3.10}) to $|p_i-k_j|\le \pi/a$. This scheme is essentially
equivalent to taking a sufficiently large cavity (for a given
momentum) -- two of the four integrals in Eq.(\ref{eq:E3.9}) become
delta--functions, and the other two can be written as
\be 4\;{\sin^2 (k_1-k_2)a \over (k_1-k_2)^2}\label{eq:E3.11}\ee
This factor is strongly peaked at $k_1=k_2$, with a corresponding
suppression of the satellites; physically, for reasonable $a$, the
detected particles can be traced back to their parent levels in the
closed cavity, and no interference is possible. For $k_1=k_2$, we
recover the two particle normalization.

\bigskip
\bigskip
\noindent
{3.2 \underline{\it Toy model 2. Emission from two one-dimensional cavities}}
\bigskip

Next, we consider two cavities, defined by walls at $L\pm a$ and
$-L\pm a$.  There are now two sets of the eigenstates, defined
separately for each cavity:
\be
\phi_{ p,N}(y) &=& (2a)^{-1/2}(2k_{0})^{-1/2}
e^{-i\,p\cdot y}~,~ ~ ~ ~{\rm for}~L_N-a\le y \le L_N+a \nonumber \\
&=& 0~,~ ~ ~ ~{\rm otherwise}. \label{eq:E4.13}\ee
The index $N$ runs over the two cavities, and $L_N = \pm L$.
The spectrum of states in each cavity
is defined by Eq.(\ref{eq:E3.7}), and the field decomposition is
\be
{\hat \varphi}(y)\;=\;\sum_{p,N}
{\hat a}_{p,N}\:\phi_{p,N}(y)~ ~ ~,\label{eq:E3.13}\ee
where the Fock operators have acquired an additional index enumerating
the cavities. States belonging to different cavities are independent, as
implied by the commutation relations:
\be
[{\hat a}_{p,N},\, {\hat a}^\dagger_{p^\prime,N^\prime} ]\;=\;
\delta_{pp^\prime}\delta_{NN^\prime}~ ~ ~.\label{eq:E3.15}\ee
With the agreement that there is a one-to-one correspondence between the
states in any one cavity and states in which we detect the particles,
the one-particle distribution reads
\be
\langle {\hat N}_k\rangle \;=\;\sum_{N}n(k,N)\;=\;\sum_{N} \: \langle
{\hat a}^\dagger_{k,N} {\hat a}_{k,N} \rangle ~ ~ ~.\label{eq:E3.16}\ee

We now consider the two-particle spectrum, and our intention to
resolve the discrepancy between Eqs.~(\ref{eq:E2.14}),
(\ref{eq:E2.24a}) and (\ref{eq:E2.24}).  In our model, there is
interference between four amplitudes: two amplitudes when each
particle is emitted from a different cavity, and two more when both
particles originate from the same cavity.  The general expression for
the two-particle spectrum reads
\be
\langle {\hat N}_{k_1}({\hat N}_{k_2}-\delta_{k_1k_2})\rangle \;=\;
\sum_{\stackrel{p_1,...,p_4}{N_1,...,N_4}}
\langle {\hat a}^\dagger_{p_1,N_1} {\hat a}^\dagger_{p_3,N_3}
{\hat a}_{p_4,N_4}
{\hat a}_{p_2,N_2}\rangle
{(\omega_{k_1}\!+\!\omega_{p_1})(\omega_{k_1}\!+\!\omega_{p_2})
(\omega_{k_2}\!+\!\omega_{p_3})(\omega_{k_2}\!+\!\omega_{p_4})
 \over 4\pi^2 4a^2 4 \omega_{k_1} \omega_{k_2}
\sqrt{\omega_{p_1}\omega_{p_2}\omega_{p_3}\omega_{p_4}} }\nonumber \\
\times\int_{L_1-a}^{L_1+a}\!\! dy\: e^{-i(k_1-p_1)y}
\int_{L_2-a}^{L_2+a}\!\! dy\: e^{i(k_1-p_2)y}
\int_{L_3-a}^{L_3+a}\!\! dy\: e^{-i(k_2-p_3)y}
\int_{L_4-a}^{L_4+a}\!\! dy\: e^{i(k_2-p_4)y}~,\label{eq:E3.17}
\ee
where $L_j=L(N_j)=\pm L$, and momentum $p_j$ originates from the cavity $N_j$.
The statistical average may be found using (\ref{eq:E3.15}):
\be
\langle {\hat a}^\dagger_{p_1,N_1} {\hat a}^\dagger_{p_3,N_3}
{\hat a}_{p_4,N_4} {\hat a}_{p_2,N_2} \rangle
\;=\;n(p_1,N_1)n(p_3,N_3)\:
\bigg [\delta_{p_1p_2}\delta_{N_1N_2}\delta_{p_3 p_4}
\delta_{N_3 N_4} +
\delta_{p_1p_4}\delta_{N_1N_4}\delta_{p_3 p_2}\delta_{N_3 N_2}\bigg ]
{}~.\label{eq:E3.18}\ee
Eq.~(\ref{eq:E3.17}) becomes
\be
\langle {\hat N}_{k_1}({\hat N}_{k_2}-\delta_{k_1k_2})\rangle \;=\;
\langle {\hat N}_{k_1}\rangle\; \langle {\hat N}_{k_2}\rangle\;
+\; \sum_{\stackrel{p_1,p_3}{N_1,N_3}} n(p_1,N_1)n(p_3,N_3)\;
{(\omega_{k_1}\!+\!\omega_{p_1})(\omega_{k_1}\!+\!\omega_{p_3})
(\omega_{k_2}\!+\!\omega_{p_1})(\omega_{k_2}\!+\!\omega_{p_3})
\over 4\pi^2 4a^2 4 \omega_{k_1} \omega_{k_2}
\omega_{p_1}\omega_{p_3} } \nonumber \\
\times {\sin (p_1-k_1)a \over (p_1-k_1)}\,{\sin (p_1-k_2)a \over (p_1-k_2)}
\,{\sin (p_3-k_1)a \over (p_3-k_1)}\,{\sin (p_3-k_2)a \over (p_3-k_2)}
\:e^{-i(L_1-L_3)(k_1-k_3)}~.\hspace*{0.7cm}\label{eq:E3.19}
\ee
The two terms in the sum with
$N_1\neq N_3$ ({\it i.e.,} $L_1\neq L_3$) lead to the usual interference
term, with periods of oscillation  $\Delta k \sim 1/(2L)$.

The resolution of the detectors
should not allow us to determine which cavity the particle originated
from, {\it i.e.} we do not have
\be
|k_1-k_2|\:L\; \gg\; 1~ ~ ~.\label{eq:E3.12}
\ee
This condition corresponds to the Rayleigh criterium, {\it i.e.},
different emission points cannot be resolved sufficiently to construct
an ``optical'' image of source. Further, in most physical situations,
$L\gg a$, so that
\be
|k_1-k_2|\:a\; \ll\; 1~ ~ ~.\label{eq:E3.12a}\ee
This last inequality effectively resolves the artificial problem
discussed at the end of the last section.
The level spacing is sufficiently
large that the analyzers do not detect particles coming from
two different levels simultaneously.  In other words, a measurement of
the one-particle spectrum does not allow for a determination of the
cavity size.

We now consider the two-particle spectrum starting from its Wigner
operator representations (\ref{eq:E2.24a}), or (\ref{eq:E2.24}).  An
optimistic expectation is that the mean coordinates $R=\pm L$ will
label the positions of the cavities, while coordinates $z$ will be
responsible for the internal dynamics of each of them, even after the
operators ${\hat N}_{k}$ and ${\hat N}_{k_1}({\hat
N}_{k_2}-\delta_{k_1k_2} )$ have been averaged.  We recall that the
introduction of Wigner distributions depending on coordinates and
momenta assumes that we have foregone a full quantum description of
the particles. However, any quantum interference problem cannot even
be formulated unless both final and initial states of the particles
are described in terms of their quantum numbers.

Both forms (\ref{eq:E2.24a}) and (\ref{eq:E2.24}) of the Wigner
operator reproduce all four terms in (\ref{eq:E3.19}), after averaging
over the set of quantum states of the two cavities.  For example,
consider Eq.~(\ref{eq:E2.24a}). Substituting Eqs.~(\ref{eq:E4.13}) and
(\ref{eq:E3.13}) into Eq.~(\ref{eq:E2.22}), we obtain
\be
\langle {\hat N}_{k_1}({\hat N}_{k_2}-\delta_{k_1k_2})\rangle &=&
\sum_{\stackrel{p_1,...,p_4}{N_1,...,N_4}}
\langle {\hat a}^\dagger_{p_1,N_1} {\hat a}^\dagger_{p_3,N_3}
{\hat a}_{p_4,N_4} {\hat a}_{p_2,N_2}\rangle
\; {(\omega_{k_1}\!+\!\omega_{p_1})(\omega_{k_1}\!+\!\omega_{p_2})
(\omega_{k_2}\!+\!\omega_{p_3})(\omega_{k_2}\!+\!\omega_{p_4})
 \over 4\pi^2 4a^2 4 \omega_{k_1} \omega_{k_2}
\sqrt{\omega_{p_1}\omega_{p_2}\omega_{p_3}\omega_{p_4}} } \nonumber \\
&\times& \int \!dR_1dR_2\,dz_1dz_2\: e^{-ik_1z_1-ik_2z_2}
\; \bigg [ e^{-ip_1(R_1+{z_1\over 2})} \bigg ]_{N_1}
\bigg [ e^{ip_2(R_2+{z_2\over 2})} \bigg ]_{N_2}
\bigg [ e^{-ip_3(R_2-{z_2\over 2})} \bigg ]_{N_3}
\bigg [ e^{ip_4(R_1-{z_1\over 2})} \bigg ]_{N_4}~.\nonumber\\
\label{eq:E3.20}
\ee
The average of the product of Fock operators has four terms, given by
Eq.~(\ref{eq:E3.18}), each corresponding to a specific combination
of cavity states.  The exponentials in (\ref{eq:E3.20}) originate from
the wave functions of the cavities, which set the limits of
integration depending of the choice of cavity.  For example, if
$N_1=N_4\rightarrow -L$ and $N_2=N_3\rightarrow +L$ then the limits of
integration are
\be
-L-a< R_1+{z_1\over 2}<-L+a, ~~~L-a< R_1-{z_1\over 2}<L+a;  \nonumber\\
-L-a< R_2-{z_2\over 2}<-L+a, ~~~L-a< R_2+{z_2\over 2}<L+a~ ~ ~.\label{eq:E3.21}
\ee
If Eq.~(\ref{eq:E2.24a}) had represented only the first term of
(\ref{eq:E2.38}), then the first line of inequalities above would have
contained only the cavity coordinate $-L$, while the second line would
have contained only $+L$. This is not the case, and we see that the
coordinates $R_j$ themselves fail to enumerate the cavities, contrary
to our naive assumption.  Thus, the requirement that the initial state
be defined in terms of quantum states makes the Wigner distribution,
which was introduced in a formal manner in Sections 2.1 and 2.2, an
inconvenient tool for handling interferometry problems.

If we consider all terms in the sum over $N_i$ in
Eq.~(\ref{eq:E3.20}), we recover the two--term formula
(\ref{eq:E3.19}) which was obtained from the original representation
of the two--particle spectrum in terms of the locally defined quantum
states.

\bigskip
\bigskip
\noindent {3.3 \underline {\it Hydrodynamics and interferometry}}
\bigskip

The generalization of the two--cavity model to the freeze--out of an
extended hydrodynamic system is evident: we consider a continuous set
of decaying cells, taking into account the curvature of the freeze--out
surface $T(x)=T_c$ and the Doppler shift of the thermal spectrum of
moving cells \cite{MS,AMS}. In addition, we neglected the effects of
the multiple local emission, {\i.e.}, the possibility that two
particles are emitted from the same elementary fluid cell. For the
one-- and two--particle spectra we then have \cite{MS}
\be
k^{0}\,{{dN_{1}} \over {d{\vec k}} } \;=\;
J(k,k) ~ ~ ~,\label{eq:E3.22}\ee
and
\be
k^{0}_{1} k^{0}_{2}\,{{dN_{2}} \over {d{\vec k}_{1} d{\vec k}_{2} }
}\; =\;   J(k_{1},k_{1})\, J(k_{2},k_{2})\:+\: {\rm Re}\bigg
[J(k_{1},k_{2})\,J(k_{2},k_{1})\bigg ]~ ~ ~,\label{eq:E3.23}\ee
respectively,
where the emission function $J(k_1,k_2)$ is given by
\be
J(k_1,k_2)\;=\;\int_{\Sigma_{c}} d\Sigma_{\mu}(x)\:{k^{\mu}_{1}+
k^{\mu}_{2}\over{2}}\: n(k_{1}\cdot u(x))\: e^{-i(k_{1}-k_{2})x}~ ~ ~.
\label{eq:E3.24}\ee
In Ref.~\cite{AMS}, these equations were used to study interferometry
for several types of one--dimensional flow.

\renewcommand{\theequation}{4.\arabic{equation}}
\setcounter{equation}{0}

\bigskip
\bigskip
\noindent {\large {\bf 4.~One--dimensional interferometry}}
\bigskip

We shall analyze in this Section only the scale--invariant
one--dimensional hydrodynamic regime, and neglect the boundary effects
which were examined in Ref.~\cite{AMS}.  In the central rapidity
region, the scale--invariant solution does not differ much from the
Landau model, provided that freeze--out takes place after sufficiently
long evolution, so that the initial longitudinal size of the
Lorentz--contracted nuclei in the c.m.s. is negligible.

First, some notation: The parameters of the model are
the critical temperature, $T_{c}~ (\sim m_{\pi})$,
and the space--like freeze--out hypersurface, defined by
$t^{2}-x^{2}_{\sst \|} = \tau^{2} = {\rm const}$.
The rapidity $y$ of a fluid cell is restricted to $\pm Y$ in the c.m.s.
We assume a Gaussian transverse distribution of hot matter in a pipe with
the effective area $S_{\bot} = \pi R^{2}_{\bot}$.
The particles are described by their momenta
\begin{equation}
 k^{\mu}_{i} = (k^{0}_{i},k^{\sst \|}_{i},{\vec p}_{i}) \equiv
(m_{i}\cosh\theta_{i},\
 m_{i}\sinh\theta_{i},\ {\vec p}_{i})
\label{eq:E4.1} \end{equation}
where ${\vec p}_{i}$ is the transverse momentum, $\theta_{i}$ the
particle rapidity in
$x_{\sst \|}$-direction, and $m^2_i = m^{2}_{\pi}+{\vec p}\ ^{2}_{i}$
is the transverse mass. Let
\be
2\alpha \;=\; \theta_1\:-\:\theta_2~,\;\; ~ ~ ~
2\theta \;=\; \theta_{1}\:+\:\theta_{2} \label{eq:E4.2}\ee
be the difference and the sum of particle rapidity, and
${\vec q}_{\bot} = {\vec p}_{1} - {\vec p}_{2}$.
The one--particle distribution has the form of
a Bose thermal distribution of pions at temperature $T=T_c$.
Since the $m_\bot$ values of interest are larger than
$T_c$ we may approximate the Bose distribution by a Boltzmann form,
and use the saddle point method to estimate the integrals
(\ref{eq:E3.24}). For the one-particle distribution these two steps yield
\be
{{dN}\over{d\theta_1 d{\vec p}_1}} \;\approx\; \tau S_\bot m_1
 \int^{Y}_{-Y} \!dy\ \cosh(\theta_1-y)
\: e^{-m_1\cosh(\theta_1-y)/T_c} \;\approx\; \tau S_{\bot}m_1
\: \sqrt{2\pi T_{c}\over{m_{1}}} \: e^{-m_{1}/T_c}~ ~ ~.
\label{eq:E4.3} \ee
The general expression for $J(k_1,k_2)$ is
\be
J(k_1,k_2) \;=\; \frac 12\: \tau S_\perp
\: e^{-{\vec q}^{\, 2}_\perp R^2_\perp/2}
\; \int^{Y}_{-Y}\!dy\ \bigg [(m_1+m_2)\,
\cosh(\theta\!-\!y)\,\cosh\alpha\: -
\: (m_{1}-m_{2})\,\sinh(y\!-\!\theta)\,\sinh\alpha \bigg
] \\ \times
\exp \bigg \{ -{1\over{T_c}} \bigg [(m_1\,+\,iF(m_1\!-\!m_2))
  \,\cosh(y-\theta)\,\cosh\alpha \:-\:(m_1\,+\,iF(m_1\!+\!m_2))\,
\sinh(y\!-\!\theta)\,\sinh\alpha\bigg ] \bigg \}\label{eq:E4.4}
\ee
where $F=\tau T_c$. After some tedious algebra, using
the saddle--point approximation, we obtain
\be
R(k_{1},k_{2}) &=& {\rm Re}\,
\bigg [J(k_{1},k_{2})J(k_{2},k_{1}) \bigg ] \nonumber \\
&=&  \frac 14 \,2\pi \mu \tau^2 S^2_\bot T_c
\; e^{-{\vec q}^{\,2}_{\bot}R^{2}_{\bot}}
\; {{g(z)\,g({1\over{z}})} \over {[h(z)\,h({1\over{z}})]^{3/2}}}
\nonumber \\ &\times& \exp \bigg \{
-{\mu \over {T_{c}}}\: \bigg [h(z)\cos {H(z)\over{2}}+h({1\over{z}})
\cos {{H({1\over{z}})}\over{2}}\bigg ] \bigg \} \nonumber \\
&\times& \cos\bigg \{
{\mu \over   {T_{c}}}\bigg [h(z)\sin {H(z)\over{2}}+h({1\over{z}})
\sin {{H({1\over{z}})}\over{2}} \bigg ]
+ {3\over 4}\bigg [H(z)+H({1\over {z}})\bigg ]
+G(z)+G({1\over{z}}) \bigg \} \nonumber \\
\label{eq:E4.5}\ee
where
$\mu = (m_{1}m_{2})^{1/2}$ and $z = (m_{1}/m_{2})^{1/2} $, and we have
introduced the functions
\be
  h(z) &=& \bigg \{ \bigg [
z^{2}-F^{2}(z-{1\over{z}})^{2}+4F^2\sinh^{2}\alpha \bigg ]^2
+4F^2(z^2-\cosh 2\alpha)^2 \bigg \}^{1/4}~;\nonumber\\
\tan H(z) &=& {2F(\cosh 2\alpha-z^{2}) \over
{z^{2}-F^{2}(z-{1\over {z}})^{2}+4F^{2}\sinh^{2}\alpha}}~;
\nonumber \\
  g(z) &=& \bigg [(z^{2}+\cosh 2\alpha)^{2}
  +F^{2}(z^{2}-{1\over{z^{2}}})^{2}\bigg ]^{1/2}~;\nonumber\\
\tan G(z) &=& {F(z^{2}-{1\over{z^{2}}}) \over {z^2+\cosh 2\alpha}}
\label{eq:E4.6}
\ee

The formulae (\ref{eq:E4.5}) and (\ref{eq:E4.6}) are useful for a
measurement of the longitudinal size in the case that the pions have
unequal transverse momenta; the case of equal transverse masses
$m_1=m_2\equiv m_\bot$ was derived in Ref.~\cite{AMS}.  The main
result in that study was that the full longitudinal size of the
freeze--out domain is not seen in the correlation function. Since 1--d
expansion typically has large velocity gradients, the local thermal
spectra of the slices with different rapidities do not overlap.  The
correlator measures the effective size of the fluid slice which forms
the observable spectrum at a given rapidity $\theta$ (or longitudinal
momentum $k^{\sst \|}$)
\cite{AMS}:
\be
\Delta y \;=\;\sqrt{{T_c \over m_{\bot}}}, ~ ~ ~ ~{\rm or}~ ~ ~ ~
\Delta x_{\sst \|} \;=\;{\tau \over {\rm cosh}\theta}\:
\sqrt{{T_c \over m_{\bot}}}~ ~ ~.\label{eq:E4.7}
\ee
These results seem to be in agreement with recent data
\cite{NA35,NA44}, where the $m_{\bot}$--dependence of the effective
longitudinal size was examined. The physical origin of the dependence
(\ref{eq:E4.7}) is very simple: the larger the transverse momentum of
a particle in the fluid, the more it is frozen into collective
longitudinal motion, and the less the spectral pattern is broadened by
thermal motion. Later we shall discuss what kind of parameters replace
$m_\perp$ in the case of transverse hydrodynamic motion.

The details of the correlator (\ref{eq:E3.23}), in particular whether
or not oscillatory behavior around unity may occur, have caused some
controversy \cite{Weiner,Ulrich}.
 The correlator derived in the formalism of the Wigner
distributions (see Eq.~(\ref{eq:E2.38})) does not reveal this
behavior, while Eq.~(\ref{eq:E3.23}) permits such oscillations. We
have already discussed why a naive usage of the Wigner formalism may
distort interference effects. There is a simple argument why
oscillations are unavoidable in this class of hydrodynamic models:
mathematically, only very restrictive conditions on the distribution
functions will yield a positive definite result.  Physically, when the
velocity gradients are large (corresponding to small $\tau$ in 1--d
expansion), two fluid cells with large relative velocity may still be
close in coordinate space, even though the bulk of their spectra are
Doppler--separated.  We then effectively have two point--like sources,
individually not resolvable, and the oscillations in the correlator
are of the same origin as in HBT interferometry of a double stellar
source.

Up to now, we have assumed an instantaneous common proper time of
emission for all particle types. However, it has been  conjectured
\cite{destil} that the hot expanding system may undergo a so--called
``strangeness distillation,'' {\it i.e.} a premature escape of the
kaons from the hadronic gas because the $K-\pi$ cross--section is
relatively small. We now discuss this phenomenon, and study how
finite emission intervals may influence the interferometric data.

A picture of sharp common freeze--out for all particles leads to the
conclusion that $\gamma
\equiv a_{\sst eff} m_\perp^{1/2}$ is independent of $m_\perp$, and
that $\gamma_\pi=\gamma_K$. This is indeed observed experimentally
\cite{NA44}, suggesting that the kaon fluid remains coupled to the
pions right until their common (sharp) freeze--out. If we allow for a
dynamical decoupling of the kaons from the pionic fluid, then the full
kaon spectra will result from a gradual emission of the kaons over some
interval $\tau_0 < \tau < \tau_\pi$. For simplicity, we shall assume
that the emission function is given by the unweighted average
\be
\langle J \rangle_\tau \;=\; \frac 1{\tau_\pi-\tau_0}\:
\int_{\tau_0}^{\tau_\pi} \! d\tau \; J(\tau)~ ~ ~,\label{eq:E4.8}
\ee
where $J$ is calculated numerically from Eq.~(\ref{eq:E3.24}),
and $\tau_\pi$ is the pion freeze--out time. In the integral
(\ref{eq:E4.8}), we assume that the temperature is a function of
$\tau$, according to $\tau T^3={\rm const}$. In subsequent
calculations we shall take $T_\pi=130~{\rm MeV}$ at $\tau_\pi=30~{\rm
fm}$.

In Fig.~\ref{salve}, we show the correlator as a function of $\Delta
k^{\sst
\|}/m_\perp^{1/2}$. The various curves correspond to different values
of $m_\perp$. The curves marked ``s'' are the result if we take
$\tau_0=\tau_\pi$ for the emission function. They coincide almost
perfectly with each other, in agreement with Eq.~(\ref{eq:E4.7}). The
curves marked ``g'' correspond to a gradual emission with
$\tau_0=7~{\rm fm}$ ($T_0=180~{\rm MeV}$).  Clearly, the
$m_\perp^{1/2}$--scaling is violated for this type of emission.
NA44 data gives no indication of such scaling violation. A
systematic and parallel study of pion and kaon interferometric source
sizes as a function of $m_\perp$ will provide a sensitive test for
strangeness distillation in hadronic matter at RHIC.

\renewcommand{\theequation}{5.\arabic{equation}}
\setcounter{equation}{0}

\bigskip
\bigskip
\noindent {\large {\bf 5.~Two-- and three--dimensional interferometry}}
\bigskip

In this Section, we discuss the interferometric measurements for two--
and three--dimensional hydrodynamic flow. To motivate this discussion,
we begin by recalling that the most important lesson from the case of
a one--dimensionally expanding source is that there is no universal
prescription for decoding interferometric data.  Thus, in interpreting
this data, we must first determine which model is applicable to the
data. The model should depend on a minimal number of parameters, and
exhibit some specific and clear behavior that is qualitatively
recognizable in the data.  Only then may we hope to extract these
unknown parameters from interferometry.  For example, in 1--d
hydrodynamic motion of the Bjorken or Landau type, the signatures are
the plateau in the central rapidity region, an increase of the
correlator width with increasing particle pair momentum, independence
of the correlator width upon the sum of the rapidities, independence
of the product $a_{eff}\sqrt{m_{\bot}}$ on the $m_{\bot}$, and an
increase of the correlation function at rapidities that exceeding the
plateau width.  If such behavior is found, one may use the
corresponding formulae and appropriate variables in fitting the data,
and, using the full dynamical description of the evolution, determine
parameters of the expanding system. An example of such a strategy was
described in Ref.~\cite{AMS}.  The abundance and clarity of signatures for
a boost--invariant--like scenario are the result of the homogeneity of
the flow parameters along the expanding hot pipe.  The contributions
from all the slices add to intensify the signal.  Mathematically, this
type of flow allows for rather simple analytic formulae, and since we are
free to choose sufficiently large $m_{\bot}$, the saddle point
approximation is very accurate.  This fact is important, because only
with analytic formulae can one obtain an understanding of the
dependence of the correlator on the physical parameters.

Unfortunately, this attractive feature of the boost--invariant
solution for one--dimensional flow is absent in two-- and
three--dimensional flows, even if the radial motion is very strong.
For radial flow, the angular dependence of the Cartesian velocity
components is weak, and the localization of the momentum spectrum is
far less than in one--dimensional expansion. These small gradients
mean that a saddle point integration will be a bad approximation, and,
consequently, we have no simple dependence of the correlator upon the
flow parameters. While even moderate radial flow does obscure the true
transverse source size, the spectrum is not sufficiently localized to
allow one to obtain a simple formula for the correlator (at least in
terms of standard variables).

Let us illustrate this using two unphysical, but exactly calculable
examples.  The first is a scale--invariant two--dimensional expansion
in the absence of longitudinal flow, as in the explosion of a long,
thin filament.  The second example is a scale--invariant
three--dimensional expansion, corresponding to a point--like explosion.
An advantage of these models is that the hydrodynamic equations are
exactly solvable, and we are guaranteed dynamical consistency
between the velocity field and the shape of freeze--out surface. This
is an important requirement for any model calculation.

\bigskip
\bigskip
\noindent {5.1 \underline{\it The explosion of a long filament}}
\bigskip

We consider the transverse expansion of a filament of length $L$, much
larger than the transverse freeze--out radius.  For this case of
purely cylindrical expansion, a convenient parametrization of the
coordinates is
\begin{equation}
 x^{\mu} \;=\; (\tau\, \cosh\beta,\: \tau\,\sinh\beta\,\cos\psi,\:
\tau\,\sinh\beta\,\sin\psi,\: z)~ ~ ~, \label{eq:E5.1}\end{equation}
while the temperature and the velocity fields may be written as
\be
\tau^2\: T^3 &=& {\rm const} \label{eq:E5.2} \\
 u^{\mu} &=& (\cosh\beta,\: \sinh\beta\,\cos\psi,\:
\sinh\beta\,\sin\psi,\: 0)~ ~ ~, \label{eq:E5.3} \ee
where $\beta$ turns out to be the radial rapidity of the fluid element.
An element of the freeze--out hypersurface is given by
$d\Sigma^{\mu} \;=\; u^{\mu}\,\tau^2\,\sinh\beta\: dz\,d\beta\, d\psi$.

The emission function $J(k_1,k_2)$ may be expressed using the
integral
\be
 {\cal J} &=& \int_0^\infty\!\sinh\beta\, d\beta \int_0^{2\pi}\! d\psi
\;\exp \bigg [ -a\cosh\beta \:+\: b_1\sinh\beta\,\cos\psi\:
+\: b_2\sinh\beta\,\sin\psi \bigg ] \nonumber \\
&=& 2\pi \int_0^\infty\!\sinh\beta\, d\beta\; e^{-a\cosh\beta}\:
I_0(\sqrt{b_{1}^{2}+b_{2}^{2}}\,\sinh\beta) \;=\;
2\pi\;{e^{-\sqrt{a^2-b_{1}^{2}-b_{2}^{2}}} \over
\sqrt{a^2-b_{1}^{2}-b_{2}^{2}}}~ ~ ~,\label{eq:E5.5}
\ee
and derivatives of ${\cal J}$ with respect to its parameters.
Introducing the shorthand notation
\begin{equation}
m_z=\sqrt{m^2+k_z^2},~~~Q^2=-(k_1-k_2)^2,~~~
H=\bigg [ 1+T^2\tau^2 {Q^2\over m_z^2} -i{T\tau Q^2 \over
2m_{z}^{2}}\bigg ]~ ~ ~, \label{eq:E5.6} \end{equation}
we find
\begin{equation}
J(k_1,k_2)\;=\;{2\pi L \tau^2 m_{z}^{2}\over T}\;
{e^{-m_z H/T}\over (m_z H/T)^3}   \; \bigg [
{m_z H\over T}+1\bigg ] \: \bigg [1+{Q^2\over 8 m_{z}^{2}}
\bigg ] \;=\; J(k_2,k_1)~ ~ ~, \label{eq:E5.7} \end{equation}
which immediately gives us the one--particle spectrum of the model:
\be
k^0\, {{dN_1} \over {d{\vec k}}} \;=\; J(k,k) \;=\;
2\pi L \tau^2 T \: {e^{-m_z/T}\over (m_z/T)}\: \bigg [{m_z \over T}-1
\bigg ]~ ~ ~.  \label{eq:E5.8} \ee

Similar to one--dimensional boost invariant flow,  we obtain a plateau
for cylindrical boost invariant expansion, but now in the
radial rapidity distribution. The dependence of the spectral
density on $m_z$ is noteworthy: the localization of the spectrum due to
radial flow is more pronounced for greater $m_z$. Particles
with $m_z/T \gg 1$ are strongly frozen into collective radial flow.
For this reason we may expect the width of the correlator in the
transverse direction to be defined by $m_z$, rather than $m_{\bot}$.

The expression for the normalized two-particle spectrum is
 \begin{equation}
C(k_1,k_2)-1 \;=\; {\rm Re}\;{ {e^{-2m_z (H-1)/T}\over H^6}
\left[{ {m_zH/T +1} \over {m_z/T +1} }\right]^2
\left[1+{Q^2 \over 8 m_{z}^{2}} \right]^2 }~ ~ ~.
\label{eq:E5.9}\end{equation}
Its explicit real form is of little practical value, except to estimate
the reliability of the saddle--point approximation for the two--  and
three--dimensional flows. We shall return to this point later.
Instead, let us first approximate (\ref{eq:E5.9}) in the case of small
differences in radial rapidities of the particles.
For the sideways and outward directions we obtain:
 \begin{equation}
C(Q)-1 \;=\; e^{- T\tau^2  Q_{out}^2/m_z}
\cos\bigg ({T\tau \over 2m_{z}^{2}}Q_{out}^{2}\bigg )
\label{eq:E5.10}\end{equation}
for $Q_{side}=0$, and
\begin{equation}
C(Q)-1 \;=\; e^{- T\tau^2 Q_{side}^2/m_z}
\cos\bigg ({T\tau \over 2m_{z}^{2}}Q_{side}^{2}\bigg )
\label{eq:E5.11}\end{equation}
for $Q_{out}=0$. We then obtain the same radii for the sideways and
outward directions:
\begin{equation}
 R_{exp}=\tau\sqrt{T\over m_z},\;\;\;\; R_{cos}= {1\over
T\tau}R_{exp}~ ~ ~.
\label{eq:E5.12}\end{equation}
{}From symmetry, it is not unexpected that the radii should be the same
for sideways and outward directions.  The radius $R_{exp}$ is defined by the
exponential, and dominates the shape of the correlator for $T\tau>1$,
while $R_{cos}$, defined by the zero of the cosine function, dominates
for $T\tau<1$. Formally, the expressions (\ref{eq:E5.10}) and
(\ref{eq:E5.11}) are valid at small relative transverse rapidities of
the pions. They very much resemble the basic formula for longitudinal
expansion. However, one should keep in mind that the latter are
obtained without the saddle--point approximation. This approximation
fails in the cylindrical case because the integrand of
Eq.~(\ref{eq:E5.5}) varies only slowly with the azimuthal angle $\psi$
over the entire domain of the integration. To clarify this statement,
we have estimated numerically by how much the integral is affected if
we retain only a small $\psi$--interval about the maximum of the
integrand.  Even for unreasonably large ratios $m_z/T\sim 10$, we must
integrate over a large interval $|\psi|<\pi/3$ in order to obtain 90\%
of the exact value of $dN/d{\vec k}$.

\bigskip
\bigskip
\noindent {5.2 \underline{\it The explosion of a point--like source}}
\bigskip

We next consider the case of spherical expansion, and parameterize
the coordinates as
\begin{equation}
 x^{\mu} \;=\; \tau\:(\cosh\beta,\: \sinh\beta\,\sin\theta\,\cos\psi,\:
\sinh\beta\,\sin\theta\,\sin\psi,\: \sinh\beta\,\cos\theta )~ ~ ~,
\label{eq:E5.13}\end{equation}
and the temperature, velocity field, and
freeze--out hypersurface volume element can be written as
\begin{eqnarray}
\tau T\:\;=\; {\rm const}~
,\;\;\;\;\;  u^{\mu}\;=\; x^{\mu} /\tau ~,\;\;\;\;\;
 d\Sigma^{\mu}\;=\; u^{\mu}\, \tau^3\,  \sinh^2\beta \,\sin\theta \: d\beta
\, d\theta\, d\psi~ . \label{eq:E5.14}\end{eqnarray}

Once again, to calculate the emission function $J(k_1,k_2)$, we use
the auxiliary integral
\begin{eqnarray}
 {\cal J}=\int_{0}^{\infty}{\rm sinh}^{2}\beta d\beta
\int_{0}^{\pi} \sin\theta d\theta \int_{0}^{2\pi} d\psi
e^{-a{\rm cosh}\beta + b_3{\rm sinh}\beta\cos\theta
+{\rm sinh}\beta\sin\theta(b_1\cos\psi + b_2 \sin\psi)}= \nonumber \\
={4\pi\over \sqrt{a^2-b_{1}^{2}-b_{2}^{2}-b_{3}^{2}} }
K_1(\sqrt {a^2-b_{1}^{2}-b_{2}^{2}-b_{3}^{2}})~~, \hspace*{2cm}
\label{eq:E5.15}
\end{eqnarray}
were $K_1$ is the modified Bessel function.  It is now
straightforward to find the one-- and two--particle spectra:
\be
 k^0\: {{dN_{1}} \over {d{\vec k}} }  \;=\;
 J(k,k)\;=\;  4\pi \tau^3 T K_2({m\over T})~ ~ ~,   \label{eq:E5.16} \\
C(k_1,k_2)-1 \;=\; {\rm Re}{\left[ {K_2(H) \over H^2 K_2(m/T)}\right]^2
\left[1+{Q^2\over 8 m^{2}}\right]^2 }~ ~ ~,~ ~ ~ ~
H \;=\; \sqrt{1+T^2\tau^2 {Q^2\over m^{2}}
 -i{T\tau Q^2 \over 2m^{2}}} \nonumber \ee
We  see that spherically symmetric boost-invariant expansion washes out
any inhomogeneity of the local thermal spectrum. This is a distinctive
feature of the spherical model.
Approximating the correlator in the case of small
differences in radial rapidity of the particles, we obtain
\begin{equation}
C(Q)-1 \;=\; e^{- T\tau^2 Q_{out}^2/m}\:{1+Q^2/4m^2  \over
1+5T^2\tau^2 Q^2/2m^2} \;
\cos\bigg ({Q^2\over m^2}({5T\tau \over 2} +{m\tau\over 2})\bigg )~ ~ ~.
\label{eq:E5.18}\end{equation}
The radius which may be extracted from this correlator is
\begin{equation}
 R_{exp}=\tau\sqrt{T \over m}~ ~ ~,
\label{eq:E5.19}\end{equation}
and the model has no large parameter which could allow an asymptotic estimate
of the emission function.

\renewcommand{\theequation}{6.\arabic{equation}}
\setcounter{equation}{0}

\bigskip
\bigskip
\noindent {\large {\bf 6.~Conclusion}}
\bigskip

Preliminary results recently reported by the NA35 and NA44
collaborations seem to confirm the main predictions of interferometry
for a hydrodynamic one--dimensional flow scenario of the heavy ion
collision.  The intensity correlations provide a multidimensional test
of the source, and the fact that so many parameters coincide can
hardly be accidental.  Thus, we have strong evidence that even at
SPS energies ($s^{1/2} \sim 10~{\rm A\cdot GeV}$), a hydrodynamic
regime develops, and that the freeze--out takes place during a short
interval.  This collective behavior can be expected to occur at RHIC
energies, and thus it is highly desirable to continue developing the
formalism.

In this paper, we have emphasized the most important physical aspects
of the theory, analyzed the reliability of its different modifications
known in the literature, and tried to isolate some controversial
points.  We argue that, strictly speaking, interferometry does not
permit the initial data to be given in terms of semi--classical
distributions, unless this description is augmented by a clear
indication of the length scale that defines the quantum states of the
particles.  Our conclusion here is that the traditional operator
approach, based on the precise definition of the particle states,
provides a firm footing for the calculation of the two--particle
spectra. In this case, distributions such as $n(p,N)$ in
Eq.~(\ref{eq:E3.18}), and $n(k,x)$ in Eq.~(\ref{eq:E3.24}) can be
thought of as on--shell Wigner phase--space distributions of the
preceding kinetic stage.  For any type of hydrodynamic expansion, no
principal or technical problems are anticipated.

We have extended previous calculations \cite{AMS} for the
longitudinally expanding system to the case of unequal transverse
momenta, and have shown that the HBT correlator may carry information
about strangeness distillation. From the NA44 data analysis, it seems
that such distillation does not occur at SPS energies.  We further
considered transversally expanding sources, and demonstrated that in
this case one has to change the parameterization of the correlator.

We emphasize that, while HBT for hydrodynamic sources is well
understood as a physical phenomenon, it is always a problem to choose
the ``correct'' model for fitting the data. We impose the condition
that the model should allow one to recognize
it via a qualitative analysis. Only then can one hope to understand
what parameters are responsible for the correlator behavior, and find
their values by fitting the data.  Only after such an analysis has
been performed, does it become possible to use the dynamical equations
of the model, and trace the freeze--out parameters back to the earlier
stages of evolution and, eventually, to an estimate of the energy
density.  Practically, this requirement means that we must begin with
an analytic expression for a solution of the relativistic hydrodynamic
equations. This guarantees the consistency between the shape of the
freeze--out surface and the velocity field. Unfortunately, analytic
solutions for the case of three--dimensional expansion are not yet
known (the equations are nonlinear, and the initial data is not well
known).  A reasonable analytic approximation will do, but to our
knowledge, a suitable expression has not yet been derived. An
approximate formula describing a realistic, expanding system at the
freeze--out stage is an important problem for boson interferometry at
RHIC.

\bigskip
\bigskip
\noindent {\bf Acknowledgments}
\medskip

This work was supported in part by the U.S. Department of Energy
under Contract No.~DE--FG02--94ER40831.  GW acknowledges useful
discussions with Pawel Danielewicz. AM is grateful to Scott Pratt
for stimulating discussions.

\renewcommand{\theequation}{A.\arabic{equation}}
\setcounter{equation}{0}

\bigskip
\bigskip
\noindent {\large {\bf Appendix}}
\bigskip

Descriptions of nuclear collisions in terms of semi--classical kinetic
equations rely heavily on the classical nature of the source that emits the
final--state particles. Such sources may be expressed naturally in terms of
Wigner functions, and it is therefore of interest to what extent
these functions may be used directly to solve the interferometry problem.
To our knowledge, equations like (\ref{eq:E2.38}) and  (\ref{eq:E2.39})
were originally derived in a model with  classical sources. Here we revisit
this derivation.

Let $\varphi(x)$ be a quantum pion field that
is emitted by a classical current
$j(x)$. The corresponding solution of the Schr\"odinger equation for the
state vector is a coherent state $|\Phi\rangle $, with the property that
\be
\langle \Phi |{\hat \varphi}(x)| \Phi \rangle \;=\; \varphi_{cl}(x)~ ~ ~,
\label{eq:A1}
\ee
where the classical field $\varphi_{cl}(x)$ obeys the inhomogeneous  wave
equation with source $j(x)$:
\be
(\Box +m^2)\,\varphi_{cl}(x)\;=\;j(x)~ ~ ~.
\label{eq:A2}
\ee
For space--time regions outside of where the currents $j(x)$ are localized,
$\varphi_{cl}(x)$ obeys the homogeneous wave equation,
and may be expanded in the plane--wave modes of Eq.~(\ref{eq:E2.1}):
\be
\varphi_{cl}(x) \;=\; \int d^{\,3}k\: \alpha_{\vec k}\; f_{\vec k}(x)~ ~ ~.
\label{eq:A3}
\ee
The coherent state $|\Phi\rangle$ may be obtained by solving the following
equation for each mode:
\be
A_{\vec k}\,|\Phi_{\vec k}\rangle\; =\;\alpha_{\vec k}|\Phi_{\vec k}\rangle
{}~ ~ ~,\label{eq:A4}
\ee
and, after normalizing to $\langle \Phi |\Phi\rangle =1$, one obtains the
representation
\be
|\Phi_{\vec k}\rangle \;=\; e^{-|\alpha_{\vec k}|^2/2}
\sum_{n_k=0}^{\infty} {\alpha_{\vec k}^{n_k} \over (n_k !)^{1/2} } \:
|n_k\rangle \;=\;
e^{-|\alpha_{\vec k}|^2/2}
\sum_{n_k=0}^{\infty} {\alpha_{\vec k}^{n_k}A^{\dag n_k}_{\vec k}
\over n_k ! }\: |0\rangle~ ~ ~.\label{eq:A5}
\ee
{}From the equation of motion for the classical field,
(\ref{eq:A2}), one immediately obtains
\be
\alpha_{\vec k}\;=\; i
(2\pi)^{-3/2}(2\omega_{k})^{-1/2}\: j(\omega_k,{\vec k})~ ~ ~,
\label{eq:A6}
\ee
so that the coherent state $|\Phi\rangle$ may be rewritten as
\be
|\Phi\rangle \;=\;
\exp\left[ -\int\! d^{\,3}k \:{|j(\omega_k,{\vec k})|^2 \over
4 \omega_{k} (2\pi)^{3} }\right]\;
\exp\left[ -i\int\! d^{\, 3}k\: {j(\omega_k,{\vec k}) \over
(2\pi)^{3/2}(2\omega_{k})^{1/2}} \: A^{\dag}_{\vec k} \right]\: |0\rangle
\label{eq:A7}
\ee
We may also write $|\Phi\rangle = S|0\rangle$, and it can be shown explicitly
that the expression (\ref{eq:A7}) emerges if the $S$--matrix
is taken in the form of a normal product.

Evaluating the one-- and two--particle spectra is now a question of
determining the right hand sides of
\be
\langle\!\langle N_{{\vec k}} \rangle\!\rangle
&=& \langle \Phi |{\hat A}^\dagger_{{\vec k}}
{\hat A}_{{\vec k}}\ |\Phi\rangle \label{eq:A8} \\
\langle\!\langle N_{{\vec k}_1}(N_{{\vec k}_2}-
\delta({\vec k}_1-{\vec k}_2) )\rangle\!\rangle
&=& \langle \Phi |{\hat A}^\dagger_{{\vec k}_{1}}
{\hat A}^\dagger_{{\vec k}_{2}}
{\hat A}_{{\vec k}_{2}}{\hat A}_{{\vec k}_{1}} |\Phi\rangle~ ~ ~,
\label{eq:A9}
\ee
where $\langle\!\langle...\rangle\!\rangle $ denotes an average over the
statistical ensemble of classical currents.  These equations may be
obtained immediately from Eqs.(\ref{eq:E2.4}) and (\ref{eq:E2.7}),
if the density matrix is taken of the form
\be
S \rho_{in} S^{\dag}\;=\; |\Phi\rangle \langle \Phi |~ ~ ~.
\label{eq:A10}
\ee

Since the coherent state $|\Phi\rangle$ is the eigenstate of every
annihilation operator ${\hat A}_{{\vec k}}$, with eigenvalue
$\alpha_{\vec k}$, we immediately obtain the final result
\be
\langle\!\langle N_{{\vec k}} \rangle\!\rangle &=&
\langle\!\langle \alpha^\ast_{{\vec k}}
\alpha_{{\vec k}} \rangle\!\rangle \;=\;
\langle\!\langle {j^\ast(\omega_k,{\vec k})
j(\omega_k,{\vec k})\over 2\omega_k (2\pi)^3} \rangle\!\rangle
\label{eq:A11}\\
\langle\!\langle  N_{{\vec k}_1}(N_{{\vec k}_2}-
\delta({\vec k}_1-{\vec k}_2) )\rangle\!\rangle
&=& \langle\!\langle \alpha^{\ast}_{{\vec k}_1}
\alpha^{\ast}_{{\vec k}_2} \alpha_{{\vec k}_1}
\alpha_{{\vec k}_2}   \rangle\!\rangle
\;=\; \langle\!\langle { j^{\ast}(\omega_{k_1},{\vec k}_1)
j^{\ast}(\omega_{k_2},{\vec k}_2)
j(\omega_{k_1},{\vec k}_1) j(\omega_{k_2},{\vec k}_2)
\over 4 \omega_{k_1} \omega_{k_2} (2\pi)^6 } \rangle\!\rangle
\label{eq:A12}
\ee
The Fourier transform of the currents is defined in a standard way:
\be
j(k)\;=\; j(k_0,{\vec k}\,)\;=\;\int d^4 x j(x) e^{ikx}~ ~ ~,\label{eq:A13}
\ee
and we therefore obtain our previous answer for the one--particle spectrum
\be
 \langle\!\langle N_{\vec k} \rangle\!\rangle=
\int {d^{\, 4}xd^{\, 4}y\over 2\omega_k (2\pi)^3}\; e^{-i\,k\cdot (x-y)}
\;  \langle\!\langle j^{\ast} (x) \: j(y) \rangle\!\rangle ~ ~ ~.
\label{eq:A14}\ee
The two--particle spectrum now reads
\be
\langle\!\langle  N_{{\vec k}_1}(N_{{\vec k}_2}-
\delta({\vec k}_1-{\vec k}_2) )\rangle\!\rangle \;=\;
\int\! {d^{\,4}x\, d^{\,4}y\, d^{\,4}x^\prime \, d^{\,4}y^\prime
\over 4\omega_{k_1}\omega_{k_2} (2\pi)^6}\; e^{-i\,k_1\cdot
(x-y)-i\,k_2\cdot (x^\prime-y^\prime)}
\;  \langle\!\langle j^\ast(x)\, j^\ast(x^\prime)\,
j(y)\, j(y^\prime) \rangle\!\rangle ~ ~ ~,
\label{eq:A15}
\ee
{\it i.e.}, we recover the well known result that a coherent source
does not lead to any {\it quantum} interference effects. In other
words, quantum evolution follows a single trajectory, which ends up as
a pure coherent state of the pion field. After we have averaged over
this state, further consideration of the two--pion amplitudes and
their interference is impossible. However, we note that there remains
the opportunity to account for statistical effects inherent in the
distribution of the classical currents. An example is given in
Ref.~\cite{Mak72}: coherent light propagating in a fluid induces
(classical) polarization currents in the molecules. The latter become
sources for a secondary (scattered) field. Intensity correlations
of this field allow one to study the multi-particle distributions in
the fluid. For simple fluids, the exact solution of the inverse
problem is possible \cite{Mak72}.

One may formally consider the $S$-matrix in Eq.(\ref{eq:A7}) as  the
perturbation series (assuming, {\it e.g.}, that the currents are weak).
For such an expansion it is convenient to use the $S$-matrix in
the form of a $T$-ordered exponent:
\be
S\;=\; T \exp \bigg \{ -i \int d^4 x [j^{\ast}(x){\hat \varphi}(x) +
j(x){\hat \varphi}^{\dag}(x)]\bigg \}~ ~ ~.\label{eq:A16}
\ee
If only terms up to second order in the currents
are retained, then the coherent state is effectively replaced by
a two-particle state. This state is no longer the eigenstate of the
annihilation operator, and the matrix element that must be evaluated
now reads
\be
\int\! d^{\,4}x\, d^{\,4}y\, d^{\,4}x^\prime \, d^{\,4}y^\prime
\;  \langle\!\langle j^\ast(x)\, j^\ast(x^\prime)\,
j(y)\, j(y^\prime) \rangle\!\rangle \;
\langle 0| {\hat \varphi}(x){\hat \varphi}(x^\prime)
{\hat A}^\dagger_{{\vec k}_{1}} {\hat A}^\dagger_{{\vec k}_{2}}
{\hat A}_{{\vec k}_{2}}{\hat A}_{{\vec k}_{1}}
 {\hat \varphi}^{\dag}(y){\hat \varphi}^{\dag}(y^\prime) |0\rangle
\label{eq:A17}
\ee
Commuting the creation and annihilation operators
with the field operators we arrive at
\be
\int\! d^{\,4}x\, d^{\,4}y\, d^{\,4}x^\prime \, d^{\,4}y^\prime
\langle\!\langle j^\ast(x)\, j^\ast(x^\prime)\,
j(y)\, j(y^\prime) \rangle\!\rangle \;
 { [e^{ik_1x+ik_2x'}+ e^{ik_1x'+ik_2x} ][e^{ik_1y+ik_2y'}+ e^{ik_1y'+ik_2y} ]
\over   4\omega_{k_1}\omega_{k_2} (2\pi)^6}~ ~ ~.
\label{eq:A18}
\ee
Choosing instead of $(x,x')$ the coordinates $(R,z)$, we obtain
the form
\be
4 k_1^0 k_2^0 {dN\over d{\vec k}_1 d{\vec k}_2}\;=\;
\int d^{\,4}R_1 d^{\,4}R_2 \;
\bigg [ F(R_1,k_1)F(R_2,k_2)\: +\:
 e^{-i\,(R_1-R_2)\cdot (k_1-k_2)}\: F(R_1,{k_1+k_2 \over 2})
F(R_2,{k_1+k_2 \over 2}) \bigg ]~, \label{eq:A19}
\ee
where
\be
F(R,p)\;=\;\int {d^{\,4}z\over (2\pi)^4}\; e^{-i\,p\cdot z}\;
j^\ast(R+z/2)\: j(R-z/2) \label{eq:A20}\ee
is the Wigner representation of the current product. In calculating
the average $\langle\!\langle...\rangle\!\rangle$, one usually assumes
that the product of four currents factorizes into the binary products.

Clearly, this way of reasoning contradicts the concept of the
classical current itself: if the field is classical, then the number
of quanta is undefined, in principle. The selection of fluctuations
with the exactly two emitted quanta is an additional measurement of
the intermediate state of the system.  This measurement prepares a new
state of the system and removes all information about the classical
nature of the emitting currents. The further evolution can now follow
two different trajectories.  This is exactly why Eq.~(\ref{eq:A18})
has acquired a typical interference structure, even before the
particles histories were traced back to their sources.

On the other hand, if we can select a state with two quanta in the
emission field, then the current is essentially a quantum mechanical
transition current.  In this case we must take into account an
explicit backward reaction in the emitting system, and we arrive
naturally at the picture which has been already considered in
connection with photon emission.  We conclude that there is no
consistent way to express the correlations in terms of off--shell
Wigner functions.

\bigskip

\noindent $^{*}$ E-mail: makhlin, gene, and
welke@rhic.physics.wayne.edu, respectively.

\bigskip
\bigskip
\bigskip
\noindent {\bf Figure Caption}
\begin{figure}
\caption{The correlation function $C_2$, as a function of
the rescaled longitudinal momentum difference, for different values of
$m_\perp$. The curves marked ``s'' correspond to sharp emission, and
those marked ``g'' are for gradual emission.}
\label{salve}
\end{figure}

\end{document}